\documentclass[aps,pra,twocolumn,superscriptaddress]{revtex4-2}
\usepackage{graphicx} 
\usepackage{amsmath}
\usepackage{braket}
\usepackage{microtype}
\usepackage{amssymb}
\usepackage{booktabs}
\usepackage{xcolor}

\begin{document}

\title{Generative Models using Multi-layer Quantum Extreme Learning Machines}
\author{Amir Azzam}
\email{ameer.azzam@qilimanjaro.tech}
\affiliation{Qilimanjaro Quantum Tech, Carrer de Vene\c{c}uela 74, 08019 Barcelona, Spain}
\affiliation{Institut de Física d’Altes Energies, IFAE, Campus UAB, Bellaterra, 08193, Barcelona}
\author{M. Pilar Casado}
\affiliation{Department de Física, UAB, Campus UAB, Bellaterra, 08193, Barcelona, Spain}
\affiliation{Institut de Física d’Altes Energies, IFAE, Campus UAB, Bellaterra, 08193, Barcelona}
\author{Michele Grossi}
\affiliation{European Organisation for Nuclear Research (CERN), 1211 Geneva, Switzerland}
\author{Arnau Riera}
\affiliation{Qilimanjaro Quantum Tech, Carrer de Vene\c{c}uela 74, 08019 Barcelona, Spain}
\author{Sofia Vallecorsa}
\affiliation{European Organisation for Nuclear Research (CERN), 1211 Geneva, Switzerland}

\begin{abstract}
Quantum reservoir computing (QRC) and the closely related quantum extreme learning machine (QELM) paradigm exploit the complex dynamics of an uncontrolled quantum system for machine learning, offering an alternative to variational quantum algorithms whose training is hindered by barren plateaus and costly circuit optimization. Because the quantum system itself is left fixed, learning reduces to fitting a linear readout of a set of measured observables, and no gradient is ever taken through the quantum device. In this work we repurpose this paradigm for \emph{generative} modeling: a fixed, randomly initialized multi-layer QELM transforms a scalar latent variable into a rich vector of measured expectation values, and only a simple linear readout is trained to map these to samples of a target distribution. The readout is fitted by minimizing the maximum mean discrepancy (MMD) between generated and target samples. We show that this algorithm is capable of reproducing a range of one- and two-dimensional distributions.
Moreover, encoding the class label on dedicated qubits enables conditional generation of different classes of distributions with the same linear readout. We show that this cannot be achieved by the linear readout alone, and that the reservoir dynamics play a role in the ability to distinguish the different classes of distributions.
We also show that using a simple six-qubit reservoir one is able to generate handwritten digits from the Digits and MNIST datasets with high accuracy. 
Moreover, replacing the Haar-random dynamics by time evolution under several physical Hamiltonians leaves the performance unchanged, provided the evolution spreads the input across the register. Finally, we show that a classical random Fourier feature model of the same size performs equally well: at the sizes we study, the QELM offers no advantage over its classical counterpart.

\end{abstract}

\maketitle

\section{Introduction}
\label{sec:intro}

Reservoir computing originated as a way to train recurrent neural networks without back-propagating through their recurrent dynamics: the recurrent part (the ``reservoir'') is left fixed and random, and only a linear readout of its state is trained~\cite{jaeger2001echo,maass2002realtime}. Extreme learning machines apply the same idea to feed-forward networks, fixing the hidden layer and training only the output weights \cite{huang2006extreme}. Both rely on a fixed high-dimensional nonlinear map whose outputs are rich enough that a linear readout suffices to approximate the desired task.

Quantum systems are natural candidates for such a map: the exponentially large Hilbert space and the entangling dynamics of a many-body system generate complex, nonlinear functions of the input that would be expensive to compute classically. This observation underlies quantum reservoir computing (QRC) introduced by Fujii and Nakajima~\cite{fujii2017harnessing} and quantum extreme learning machines (QELM) introduced in \cite{negoro2018machine, ghosh2019quantum}. Since then, these two paradigms have developed into a broad research program~\cite{mujal2021opportunities,ghosh2021neuromorphic,markovic2020quantum,nakajima2019boosting,martinezpena2021dynamical,innocenti2023potential,xiong2023fundamental,chen2020temporal,kornjaca2024largescale}. Their central appeal is trainability: unlike variational quantum circuits, whose gradients could vanish exponentially in the number of qubits over wide regions of parameter space (the barren-plateau phenomenon~\cite{mcclean2018barren}), a reservoir is never optimized through the quantum device. Only a classical linear readout is fitted, so no gradient is ever taken through the quantum device and the optimization is low-dimensional and inexpensive.

Most QRC/QELM work targets supervised learning (regression and classification), where the readout maps observables to labels. Quantum \emph{generative} modeling, by contrast, has largely been pursued with variational quantum circuit Born machines, in which a parameterized circuit is trained, typically against an MMD loss, so that its measurement statistics reproduce a target distribution~\cite{benedetti2019generative,liu2018differentiable}. These models inherit the training difficulties of variational circuits, which persist even with MMD losses~\cite{rudolph2024trainability}. Here we ask whether the reservoir recipe (fix the quantum part, train only a linear readout) can also drive a generative model, thereby sidestepping circuit optimization entirely.

Our construction draws on two ingredients. The first is \emph{data re-uploading}: encoding the input repeatedly through interleaved single-qubit rotations, which is known to increase the expressivity of a fixed-depth circuit and lets it represent rich functions of its input~\cite{perezsalinas2020data}. The second is a classical generative objective from the moment-matching literature: we train the readout by minimizing the maximum mean discrepancy between generated and target samples~\cite{gretton2012kernel}, the same criterion used by generative moment-matching networks~\cite{li2015generative,dziugaite2015training}. With the energy-distance (Riesz) kernel, MMD is equivalent to the energy distance between the two samples~\cite{szekely2013energy,sejdinovic2013equivalence}, giving a parameter-free, differentiable loss. The thesis we test is simple: \emph{data re-uploading together with entangling dynamics is enough to build a quantum reservoir whose linear readout can generate a range of target distributions}, with no training of the quantum system itself.

We find that the thesis holds. We study the generator in two settings, unconditional and class-conditional generation. In the unconditional setting, a single scalar latent suffices to reproduce a range of one- and two-dimensional distributions of very different shapes. In two dimensions the samples lie on a curve, which finite-shot measurement noise spreads over the support of the target.

For conditional generation, we compare three ways of passing the label to the model: as a one-hot vector appended to the readout only, on dedicated label qubits inside the reservoir, or both. With an affine readout, the label must enter the reservoir. With the one-hot route alone, the linear readout can only shift a distribution, so classes that differ in shape but share their mean and variance are generated as one and the same distribution. With label qubits, the generator separates up to ten classes in one and two dimensions, and adding the one-hot vector on top gives no further improvement. We then turn to high-dimensional targets, class-conditional generation of handwritten digits on the $8\times8$ Digits and the $28\times28$ MNIST datasets. With the output clipped to the pixel range, the generator reaches a LeNet-FID of $3.4$ on Digits and $4.1$ on MNIST, compared to $2.7$ and $2.1$ between two sets of real images.

Throughout this work, the reservoir dynamics is a Haar-random unitary. We show that replacing it by the time evolution under four typical Hamiltonian families gives the same results, provided the evolution runs long enough to spread the input across the register; shortening or removing the evolution degrades the generated images. The choice of Hamiltonian therefore does not matter as long as the reservoir scrambles its input. Finally, we discuss how the generator scales with the number of qubits and layers, and we compare it with a classical random Fourier feature model with the same readout and loss, which performs equally well: at the sizes we study, the quantum feature map offers no measurable advantage.

\subsection*{Related work}

Generative models built on a quantum reservoir or extreme learning machine, with only a classical readout trained, are still rare. The closest to ours is the generative QELM of P{\l}odzie{\'n}~\cite{plodzien2026generative}, which maps a random latent vector through a fixed unitary and a trained linear output layer to one- and two-dimensional distributions; we compare with it in Sec.~\ref{sec:discussion}. Quantum reservoirs have also been used as sequence generators: Miranda and Shaji learn and generate music with an autoregressive reservoir whose outputs are fed back as inputs~\cite{miranda2025quantum}, and Ferreira \emph{et al.}\ adapt the same architecture to generate levels for platform games~\cite{ferreira2025level}. For images, Wakaura uses a quantum reservoir as the generator of a generative adversarial network~\cite{wakaura2025qrgan}, trained against a classical discriminator on single-class handwritten digits, CIFAR-10 and Fashion-MNIST. There, each generated image is obtained from a training image fed through the reservoir, and the results are compared on training images, whereas our generator draws new samples from a scalar latent alone and is evaluated on held-out images against a classical random-feature baseline.

Related models train part of the quantum system. The quantum scrambling Born machine~\cite{plodzien2026scrambling} uses a fixed entangling unitary as a scrambler but optimizes single-qubit rotations, so its samples are measurement outcomes of a trained circuit rather than the output of a classical readout. The chaotic quantum diffusion model~\cite{tran2026chaotic} replaces the random unitaries of quantum diffusion models by chaotic Hamiltonian evolution, which is closer to analog hardware, to learn distributions of quantum states. Finally, hybrid quantum-classical latent diffusion models with variational circuits have been applied to medical images, and their authors point to quantum reservoirs as a barren-plateau-free alternative for such generative models~\cite{yeteraydeniz2025hybrid}. Compared with the expectation-value generator of Ref.~\cite{plodzien2026generative}, which passes a multi-dimensional random latent through a fixed unitary into an affine readout, our construction differs in three ways: (i) a multi-layer QELM with data re-uploading that generates from a single scalar latent; (ii) class conditioning through dedicated label qubits inside the reservoir, rather than a one-hot vector added to the readout, which can only shift the output distribution; and (iii) a protocol for class-conditional image generation, evaluated on held-out images against a tuned classical random-feature baseline. To our knowledge, no previous work has used a QELM with a linear readout for class-conditional image generation.

The paper is organized as follows. Section~\ref{sec:model} describes the model and its training. Section~\ref{sec:results} presents unconditional and conditional generation of one- and two-dimensional distributions and of images. Section~\ref{sec:ham} studies the dependence on the reservoir dynamics, replacing the Haar-random unitary by the time evolution under several Hamiltonians, and Section~\ref{sec:results-classical} compares the multi-layer QELM with a classical random Fourier feature model. Section~\ref{sec:discussion} discusses the results and their limitations.

\section{Model}
\label{sec:model}

\begin{figure}[htbp]
  \centering
  \includegraphics[width=\columnwidth]{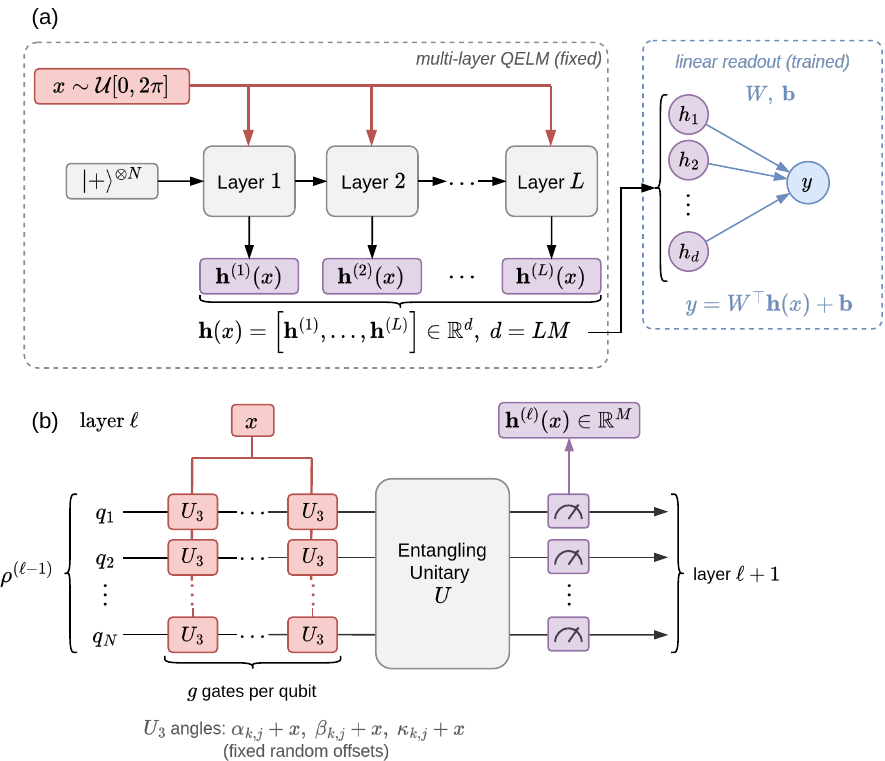}
  \caption{Structure of the multi-layer QELM generator. (a)~The latent $x\sim\mathcal{U}[0,2\pi]$ is re-uploaded into each of the $L$ layers, which act in sequence on the register prepared in $\ket{+}^{\otimes N}$. Each layer outputs $M$ expectation values $\mathbf{h}^{(\ell)}(x)$, their concatenation $\mathbf{h}(x)\in\mathbb{R}^{d}$, $d=LM$, is mapped to a sample by the linear readout, the only trained component. (b)~One layer: each qubit receives $g$ $U_3$ gates whose angles are fixed random offsets shifted by $x$, the register is propagated by a fixed Haar-random entangling unitary $U$, and $\langle Z_k\rangle$, $\langle X_k\rangle$ and $\langle Z_kZ_{k+1}\rangle$ are measured. The offsets differ between qubits and gates but are shared by all layers, and $U$ is the same in every layer. Measurement back-action is neglected, so the state passes on to layer $\ell+1$.}
  \label{diag:res}
\end{figure}

The generative model is a deterministic map from a latent variable to an output sample, built from two blocks (Fig.~\ref{diag:res}(a)): a fixed quantum feature map, the multi-layer QELM, and a trained linear readout.

The multi-layer QELM consists of $L$ layers acting in sequence on $N$ qubits. Each layer (i) encodes the latent $x$ by data re-uploading, (ii) propagates the register with a fixed entangling unitary $U$, and (iii) measures $M$ one- and two-body observables, giving the layer features $\mathbf{h}^{(\ell)}(x)\in\mathbb{R}^{M}$ (Fig.~\ref{diag:res}(b)). The register is not reset between layers: layer $\ell$ acts on the state $\rho^{(\ell-1)}$ left by layer $\ell-1$. We ignore measurement back-action, so the evolution corresponds to a restarting protocol as described in~\cite{mujal2023time}. Concatenating the $L$ layer features gives the feature vector $\mathbf{h}(x)\in\mathbb{R}^{d}$, $d=LM$, which the linear readout maps to an output sample. Only the readout is trained. The QELM stays fixed at its random initialization.

\subsection{Latent variable and input encoding}
\label{sec:encoding}

The register is initialized in the product state $\rho_0 = \big(\ket{+}\!\bra{+}\big)^{\otimes N}$, with $\ket{+} = (\ket{0}+\ket{1})/\sqrt{2}$. A single forward pass consumes one latent sample $x \sim \mathcal{U}[0, 2\pi]$, the generator's only source of randomness. In every layer, each qubit $k$ receives a sequence of $g$ single-qubit $U_3$ rotations with the latent added to all three angles:
\begin{align}
  \phi_{k,j}   &= \alpha_{k,j} + x, \\
  \gamma_{k,j} &= \beta_{k,j}  + x, \\
  \theta_{k,j} &= \kappa_{k,j} + x,
\end{align}
for $j \in \{1,\dots,g\}$. The offsets $\{\alpha_{k,j}, \beta_{k,j}, \kappa_{k,j}\}$ are drawn once, independently and uniformly from $[-2\pi, \pi]$, and then held fixed. The same encoding circuit, with the same offsets, is applied in every layer. Re-uploading the same $x$ through $g$ gates per qubit and across $L$ layers is what gives the fixed map its nonlinear dependence on the input~\cite{perezsalinas2020data}.

\begin{figure}[htbp]
  \centering
  \includegraphics[width=\linewidth]{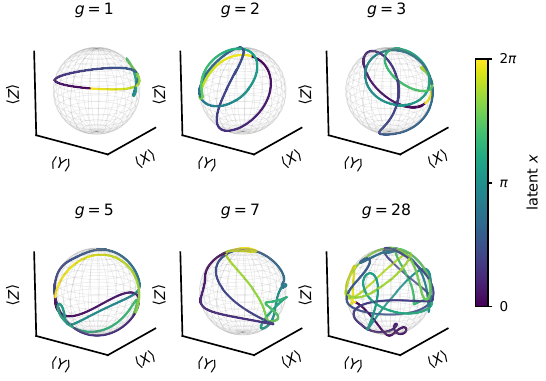}
  \caption{Bloch-sphere trajectory of a single encoded qubit as the latent $x$ sweeps $[0,2\pi]$ (color), for $g=1,2,3,5,7$ and $28$ re-uploading $U_3$ gates. With more gates the trajectory winds over a larger part of the sphere, so nearby latent variables are mapped to more distinguishable states.}
  \label{fig:bloch_map}
\end{figure}

Fig.~\ref{fig:bloch_map} shows the effect of adding more gates on the input encoding. As more encoding gates are added, the encoded input is able to span more of the Bloch sphere surface, increasing the ability to distinguish between inputs.

\subsection{Reservoir dynamics}
\label{sec:dynamics}

After encoding, the full $N$-qubit register is propagated by a fixed input-independent entangling unitary $U$, allowing information to flow between the qubits. We take $U$ to be a Haar-random unitary on the $N$-qubit register, drawn once and then held fixed. The same $U$ is applied in every layer. The Haar-random unitary is a generic, maximally scrambling choice that does not single out any particular physical system. In Sec.~\ref{sec:ham} we study the effect of replacing the Haar-random evolution with time evolution under several physical Hamiltonians and find that they provide a similar performance.

After the evolution, the expectation values of $M$ observables are measured: $\langle Z_i\rangle$ and $\langle X_i\rangle$ on every qubit and $\langle Z_iZ_{i+1}\rangle$ on nearest-neighbor pairs of an open chain, so $M = 2N + (N-1) = 3N-1$. The encoding circuit and $U$ are identical in every layer, so all layers apply the same map. Their features differ because each layer acts on a different state. In all cases we ignore the back-action of the measurement, and unless a finite number of measurement shots is stated, we use ideal (infinite-shot) expectation values. Concatenating the measurements from all layers gives the feature vector $\mathbf{h}(x) \in \mathbb{R}^d$, $d = LM$.

\subsection{Linear readout and the generator}
\label{sec:readout}

The feature vector is mapped to an output sample by a single trained linear layer,
\begin{equation}
  y = \mathbf{w}^\top \mathbf{h}(x) + b,
\end{equation}
with weights $\mathbf{w} \in \mathbb{R}^d$ and bias $b \in \mathbb{R}$. For an $n$-dimensional output the readout is $\mathbf{y} = W\mathbf{h}(x) + \mathbf{b}$, with $W\in\mathbb{R}^{n\times d}$ and $\mathbf{b}\in\mathbb{R}^{n}$. The readout is deterministic: the model's randomness comes entirely from the latent $x$. The generated distribution $q$ is therefore the pushforward of $\mathcal{U}[0,2\pi]$ under the fixed map $\mathcal{R}: x \mapsto \mathbf{h}(x)$ followed by the linear readout. Drawing a batch of latents $\{x_i\}$ and applying the map $\mathcal{R}$ yields a batch of generated samples $\{y_i\}$. 

\subsection{Training objective}
\label{sec:objective}

The readout is trained by minimizing the squared maximum mean discrepancy (MMD$^2$) between the generated distribution $q$ and the target distribution $p$~\cite{gretton2012kernel}. For a kernel $k(\cdot,\cdot)$,
\begin{equation}
  \begin{aligned}
  \mathrm{MMD}^2(p, q) ={}&
    \mathbb{E}_{t, t' \sim p}[k(t, t')]
  + \mathbb{E}_{y, y' \sim q}[k(y, y')]\\
  &- 2\,\mathbb{E}_{t \sim p,\, y \sim q}[k(t, y)],
  \end{aligned}
\end{equation}
which is non-negative and vanishes if and only if $p = q$ for a characteristic kernel. We use the Riesz (energy-distance) kernel
\begin{equation}
  k(u, v) = -\,\|u - v\|,
\end{equation}
for which MMD$^2$ reduces (up to a factor) to the energy distance between $p$ and $q$~\cite{szekely2013energy,sejdinovic2013equivalence}. Although $-\|u-v\|$ is not positive semidefinite, it is conditionally positive definite: since the weights in MMD$^2$ sum to zero, it yields, up to a factor of two, the same loss as the positive-semidefinite distance-induced kernel $k_0(u,v)=\tfrac{1}{2}(\|u\|+\|v\|-\|u-v\|)$, which is characteristic on distributions with finite first moments~\cite{sejdinovic2013equivalence}. The loss therefore vanishes if and only if $p=q$. Intuitively, the loss compares the typical distance between a generated and a target sample with the typical distance between two samples from the same set. Because the kernel is simply the Euclidean distance, it has no bandwidth or other parameter to tune, unlike the commonly used Gaussian kernel.

Given a mini-batch of $S$ generated samples $\{y_i\}_{i=1}^{S}$ and $S$ target samples $\{t_j\}_{j=1}^{S}$, the loss is the biased V-statistic estimator
\begin{equation}\label{eq:mmd_w_kernel}
  \begin{aligned}
  \widehat{\mathrm{MMD}}^2 ={}&
    \frac{1}{S^2} \sum_{i,j} k(t_i, t_j)
  + \frac{1}{S^2} \sum_{i,j} k(y_i, y_j)\\
  &- \frac{2}{S^2} \sum_{i,j} k(t_i, y_j).
  \end{aligned}
\end{equation}
Gradients propagate through the generated samples $\{y_i\}$ to the readout weights $(\mathbf{w}, b)$, which are updated using the Adam optimizer \cite{kingma2014adam}. The quantum reservoir is fixed: its encoding offsets and entangling unitary carry no gradient and are never updated.

Before training, the target samples are standardized to zero mean and unit variance, $\tilde{t} = (t - \mu)/\sigma$, where $\mu$ and $\sigma$ are the empirical mean and standard deviation of the target data, and the readout is trained to match the distribution of $\tilde{t}$. After training, the scale is folded back into the readout parameters, $\mathbf{w} \to \sigma\,\mathbf{w}$ and $b \to \sigma\, b + \mu$, so the generator remains a single affine map of the reservoir features that outputs samples directly in the units of the target.

\subsection{Conditional generation}
\label{sec:cond_pred}

\begin{figure}[htbp]
  \centering
  \includegraphics[width=\columnwidth]{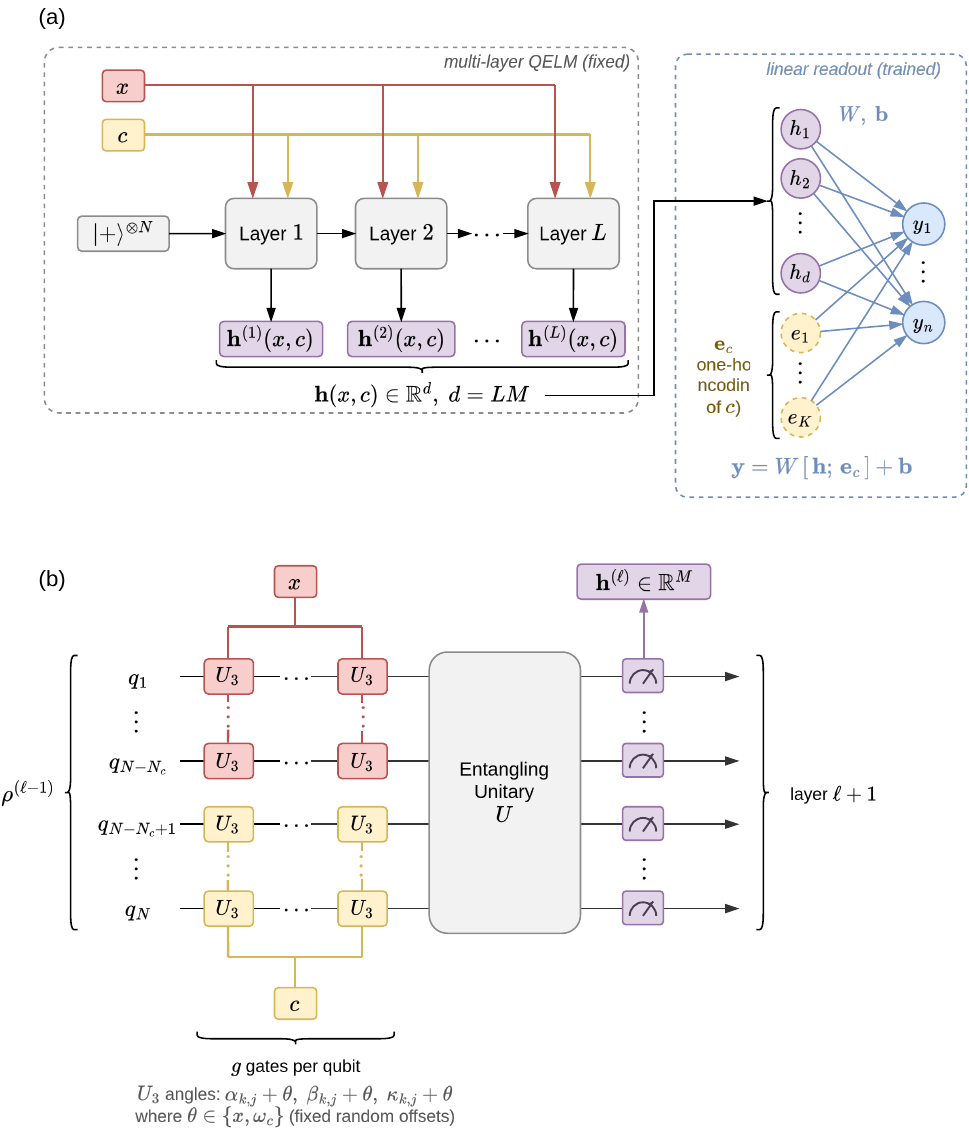}
  \caption{Structure of the conditional multi-layer QELM. (a)~The latent $x$ and the label $c$ are re-uploaded into every layer, the features $\mathbf{h}(x,c)$ are concatenated with an optional one-hot vector $\mathbf{e}_c$ (dashed) and mapped to a sample by the trained linear readout. (b)~One layer: the first $N-N_c$ qubits receive the latent $x$ and the last $N_c$ qubits receive the class angle $\omega_c=\pi c/(K-1)$, both through $g$ re-uploading $U_3$ gates whose fixed random offsets are shared by all layers, the rest of the layer is as in Fig.~\ref{diag:res}(b). In the one-hot-only variant there are no label qubits: all $N$ qubits receive $x$, as in Fig.~\ref{diag:res}.}
  \label{diag:cond_res}
\end{figure}

The model extends to class-conditional generation: a single generator is trained on $K$ target distributions $p_0,\dots,p_{K-1}$ and, given a label $c\in\{0,\dots,K-1\}$, produces samples from $p_c$. We pass the label to the model through label qubits, through a one-hot vector at the readout, or both (Fig.~\ref{diag:cond_res}).

\textbf{Label qubits.} The last $N_c$ qubits of the register carry the label and the remaining $N-N_c$ qubits carry the latent. The label enters as the angle $\omega_c=\pi c/(K-1)$, encoded by the same re-uploading gates as the latent: on a label qubit the $U_3$ angles are $\alpha_{k,j}+\omega_c$, $\beta_{k,j}+\omega_c$ and $\kappa_{k,j}+\omega_c$. The features $\mathbf{h}(x,c)$ therefore depend on the class, and the readout keeps its unconditional form, $\mathbf{y}=W\mathbf{h}(x,c)+\mathbf{b}$. In all our distribution experiments $N_c=1$, so the $K$ classes are distinguished by $K$ evenly spaced angles on a single qubit.

\textbf{One-hot readout.} The reservoir is left class-agnostic: all $N$ qubits carry the latent, exactly as in the unconditional model, and the label enters only the readout, as a one-hot vector $\mathbf{e}_c\in\{0,1\}^{K}$ concatenated with the features,
\begin{equation}
  \mathbf{y} = W\,[\,\mathbf{h}(x),\,\mathbf{e}_c\,] + \mathbf{b} = W_h\,\mathbf{h}(x) + \mathbf{w}_c + \mathbf{b},
\end{equation}
where $\mathbf{w}_c$ is the column of $W$ selected by $\mathbf{e}_c$. The label therefore adds a class-dependent constant to every sample: it can translate a class's distribution but cannot change its shape.

\textbf{Both.} The label is encoded on label qubits and also appended as a one-hot vector, $\mathbf{y}=W[\mathbf{h}(x,c),\mathbf{e}_c]+\mathbf{b}$.

In every variant the only trained parameters are those of a single linear layer, $(W,\mathbf{b})$. Training minimizes the class-averaged loss $\frac{1}{K}\sum_{c}\widehat{\mathrm{MMD}}^2_c$, where each term compares a mini-batch of 256 generated samples of class $c$ with 256 target samples from $p_c$. The targets are standardized with a mean and standard deviation pooled over all classes.

\section{Results}
\label{sec:results}

Unless stated otherwise, all results use Haar-random reservoir dynamics with $N=6$ qubits,
$g=5$ re-uploading gates per qubit and $L=7$ layers, with $M=17$ observables measured per
layer ($d=119$ features).

\subsection{One-dimensional unconditional target distributions}
\label{sec:results-1d}

\begin{figure}[htbp]
  \centering
  \includegraphics[width=\columnwidth]{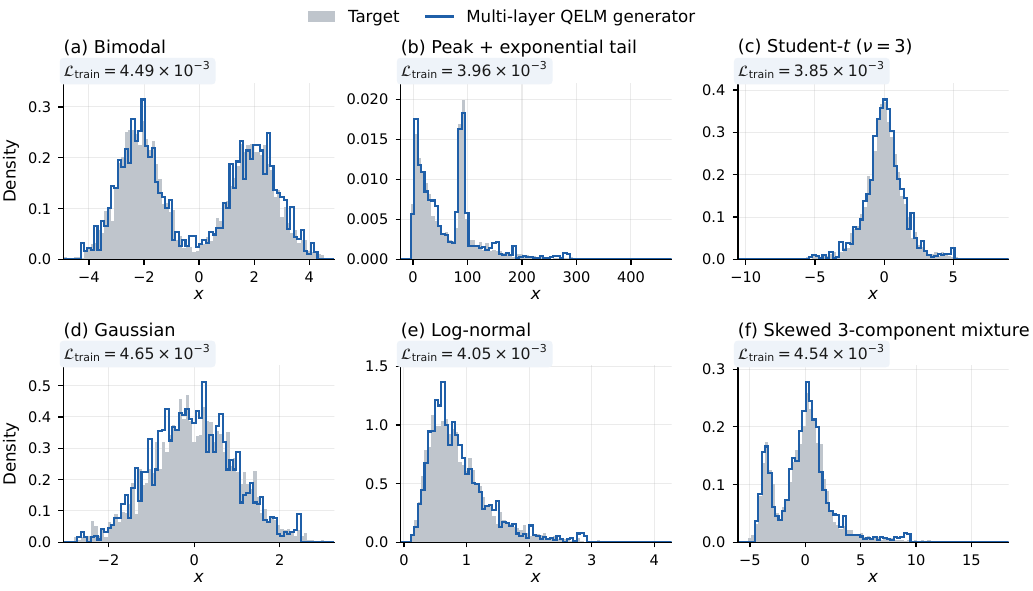}
  \caption{Six unconditioned one-dimensional target distributions learned by the multi-layer QELM generator: (a)~bimodal Gaussian mixture, (b)~peak with an exponential tail, (c)~Student-$t$ with $\nu=3$, (d)~Gaussian, (e)~log-normal and (f)~skewed three-component mixture. All panels use the same fixed Haar-random reservoir ($N=6$, $g=5$, $L=7$, $M=17$, so $d=119$) driven by a single scalar latent $x\sim\mathcal{U}[0,2\pi]$, only the linear readout $(\mathbf{w},b)$, with $d+1=120$ parameters, is trained separately for each target. Each readout is trained for 6000 Adam steps with mini-batches of $S=512$ on $\widehat{\mathrm{MMD}}^2$ with the Riesz kernel, with targets standardized to zero mean and unit variance. Gray histograms show 3000 target samples and blue outlines show 3000 generated samples. Each panel shows one training seed. $\mathcal{L}_{\mathrm{train}}$ is the final value of $\widehat{\mathrm{MMD}}^2$, averaged over the last 100 training steps.}
  \label{res:1d}
\end{figure}

Fig.~\ref{res:1d} shows the generator trained on six one-dimensional targets. With a single scalar latent, the generator is a fixed smooth function $y(x)$ of $x\sim\mathcal{U}[0,2\pi]$, and any one-dimensional distribution can be written in this form, so the task reduces to approximating one function of one variable with a linear combination of the $d=119$ reservoir features. With $d+1=120$ trained parameters the generator reproduces unimodal, multimodal, skewed and heavy-tailed targets, including the sharp peak on top of the exponential tail in Fig.~\ref{res:1d}(b). The final training loss is of order $4\times10^{-3}$ for all six targets, comparable to the step-to-step fluctuation of the mini-batch estimate, so every readout has converged.

Some systematic deviations follow from the construction. For instance, the generated samples are bounded: every feature is an expectation value of a Pauli operator, and in practice the learned map covers a narrower range than the target. For the Student-$t$ target the generated samples lie within $[-5.2,\,5.1]$, while $0.1\%$ of the target samples lie beyond $8.1$. The long tails of the log-normal, the skewed mixture and the exponential tail are cut off in the same way. 

\subsection{Two-dimensional unconditional target distributions}
\label{sec:results-2d}

\begin{figure}[htbp]
  \centering
  \includegraphics[width=\columnwidth]{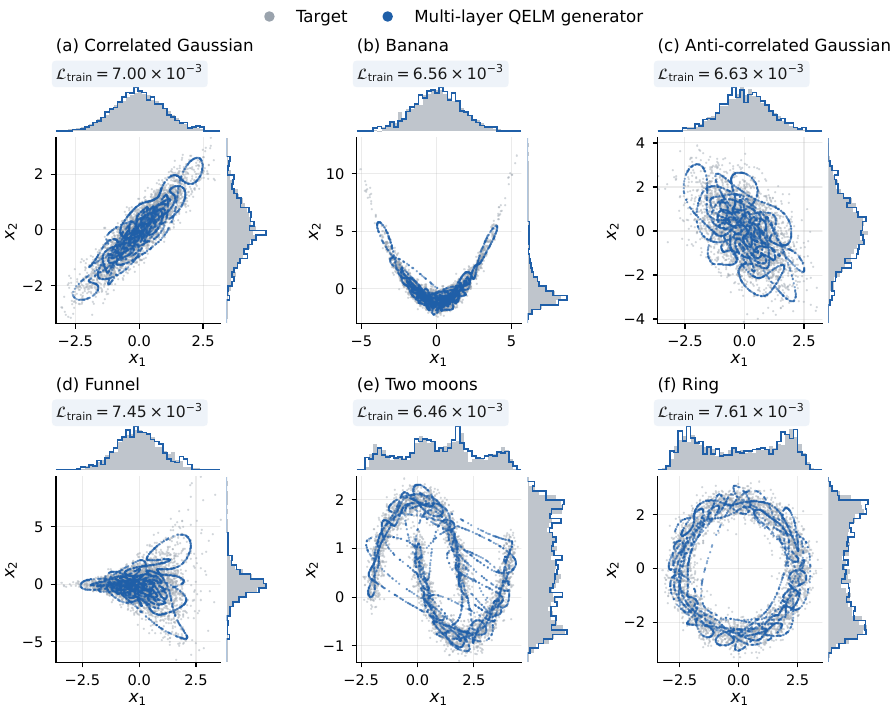}
  \caption{Six unconditioned two-dimensional target distributions learned by the multi-layer QELM generator: (a)~correlated Gaussian with unit variances and $\mathrm{Cov}(x_1,x_2)=0.9$, (b)~banana, $x_2 = 0.45\,(x_1^2-2.5)+\epsilon$, (c)~anti-correlated Gaussian with $\mathrm{Var}(x_1)=1$, $\mathrm{Var}(x_2)=1.6$ and $\mathrm{Cov}(x_1,x_2)=-0.65$, (d)~funnel, $x_2\sim\mathcal{N}(0,e^{1.2\,x_1})$, (e)~two moons and (f)~ring. All panels use the same fixed Haar-random reservoir ($N=6$, $g=5$, $L=7$, $M=17$, so $d=119$) driven by a single scalar latent $x\sim\mathcal{U}[0,2\pi]$, only the linear readout $\mathbf{y} = W\mathbf{h}(x)+\mathbf{b}$, with $W\in\mathbb{R}^{2\times d}$, $\mathbf{b}\in\mathbb{R}^2$ and $2(d+1)=240$ parameters, is trained separately for each target. Each readout is trained for 6000 Adam steps with mini-batches of $S=512$ on $\widehat{\mathrm{MMD}}^2$ with the Riesz kernel, with each target coordinate standardized to zero mean and unit variance. Each panel shows 3000 target samples (gray) and 3000 generated samples (blue) together with their marginal histograms. Each panel shows one training seed. $\mathcal{L}_{\mathrm{train}}$ is the final value of $\widehat{\mathrm{MMD}}^2$ on the standardized targets, averaged over the last 100 training steps.}
  \label{res:2d}
\end{figure}

Fig.~\ref{res:2d} shows the same generator trained on six two-dimensional targets, now with a two-component readout $\mathbf{y} = W\mathbf{h}(x) + \mathbf{b}$, where $W\in\mathbb{R}^{2\times d}$ and $\mathbf{b}\in\mathbb{R}^{2}$. The generated samples reproduce the location and the marginals of each target, but they lie on a curve rather than filling the plane. This follows directly from the construction: the generator is a deterministic, continuous map from a single scalar latent $x$ to $\mathbb{R}^2$, so its image is a one-dimensional curve, $x\mapsto\mathbf{y}(x)$, and every generated sample lies on it. Training can bend this curve so that it passes through the high-density regions of the target, which is enough to match the marginals, but it cannot give the generated distribution a non-zero area. The effect is clearest for targets with thin or disconnected support, where the curve has to cross low-density regions to connect different parts of the target, as with the chords between the two moons in Fig.~\ref{res:2d}(e).

\begin{figure}[htbp]
  \centering
  \includegraphics[width=\columnwidth]{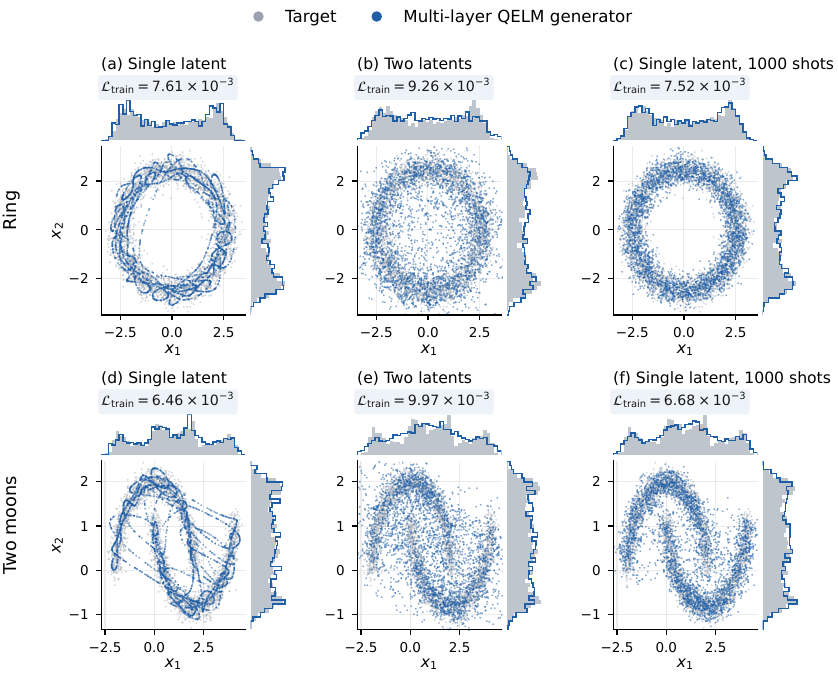}
  \caption{Line tracing of the single-latent generator and two ways to remove it, shown on the two targets where it is most visible: ring (top row) and two moons (bottom row). All panels use the same fixed Haar-random reservoir ($N=6$, $g=5$, $L=7$, $d=119$) and a linear readout trained on $\widehat{\mathrm{MMD}}^2$, only the source of randomness changes. (a,~d)~A single scalar latent $x\sim\mathcal{U}[0,2\pi]$ re-uploaded on every qubit, with ideal expectation values: the generated samples lie on a one-dimensional curve. (b,~e)~Two independent latents. (c,~f)~A single scalar latent with every expectation value $\langle O_j\rangle$ estimated from 1000 measurement shots, with the shot noise drawn independently for every training batch and every generated sample. Each panel shows 3000 target samples (gray) and 3000 generated samples (blue) with their marginal histograms, for one training seed. $\mathcal{L}_{\mathrm{train}}$ is the final value of $\widehat{\mathrm{MMD}}^2$ on the standardized targets, averaged over the last 100 training steps.}

  \label{res:2d_support}
\end{figure}

The curve can be thickened by adding a second source of randomness, and we compare two ways of doing this in Fig.~\ref{res:2d_support}, for the ring and the two moons. The first adds a second independent latent, so that the generator becomes a map from a two-dimensional latent space and its image becomes a surface. The second keeps the single latent but replaces the ideal expectation values by estimates from a finite number $N_{\mathrm{shots}}$ of measurement shots. Every observable $O_j$ has eigenvalues $\pm1$, so its estimate is $\hat{h}_j(x) = 2k_j/N_{\mathrm{shots}} - 1$, with $k_j \sim \mathrm{Binomial}\big(N_{\mathrm{shots}}, (1+h_j(x))/2\big)$ the number of $+1$ outcomes. We write $\hat{\mathbf{h}}(x)=\mathbf{h}(x)+\boldsymbol{\epsilon}$, where the shot noise $\boldsymbol{\epsilon}$ has zero mean and variance $(1 - h_j(x)^2) / N_{\mathrm{shots}}$ in component $j$, and is drawn independently for every observable and every sample. Independent binomial estimates assume a separate measurement ensemble for each observable. In practice $\langle Z_i\rangle$ and $\langle Z_iZ_{i+1}\rangle$ would be estimated from the same $Z$-basis bitstrings, and only $\langle X_i\rangle$ requires a second basis, which correlates the noise; we keep the independent model as a simple proxy. Moreover, since we ignore measurement back-action (Sec.~\ref{sec:model}), the features of layer $\ell$ are obtained in the restarting protocol by re-running layers $1,\dots,\ell$, so one sample requires $2LN_{\mathrm{shots}}$ circuit executions.

The readout then maps each sample to $\mathbf{y} = W\mathbf{h}(x) + W\boldsymbol{\epsilon} + \mathbf{b}$, which spreads the samples around the curve learned in the noiseless case, by an amount set by the number of shots. Both approaches remove the line tracing, but with 1000 shots the generated samples (Fig.~\ref{res:2d_support}(c)) follow the target more closely than with two latents (Fig.~\ref{res:2d_support}(b)). In the case of two latents, both of them pass through the same fixed scrambling dynamics, and the linear readout cannot control how far the second one moves a sample away from the curve, so samples spill into the hole of the ring and the gap between the moons. Shot noise instead perturbs a curve that already follows the target closely, so a small spread fills the support while preserving its shape. The amount of noise matters: too few shots blur the target, and too many leave the curve visible again. This approach requires no change to the architecture: the readout is trained on noisy features, with the shot noise redrawn for every training batch, and finite-shot noise is in any case unavoidable when the observables are measured on hardware.


\subsection{One-dimensional conditional target distributions}
\label{sec:1d_cond}

\begin{figure}[htbp]
  \centering
  \includegraphics[width=\columnwidth]{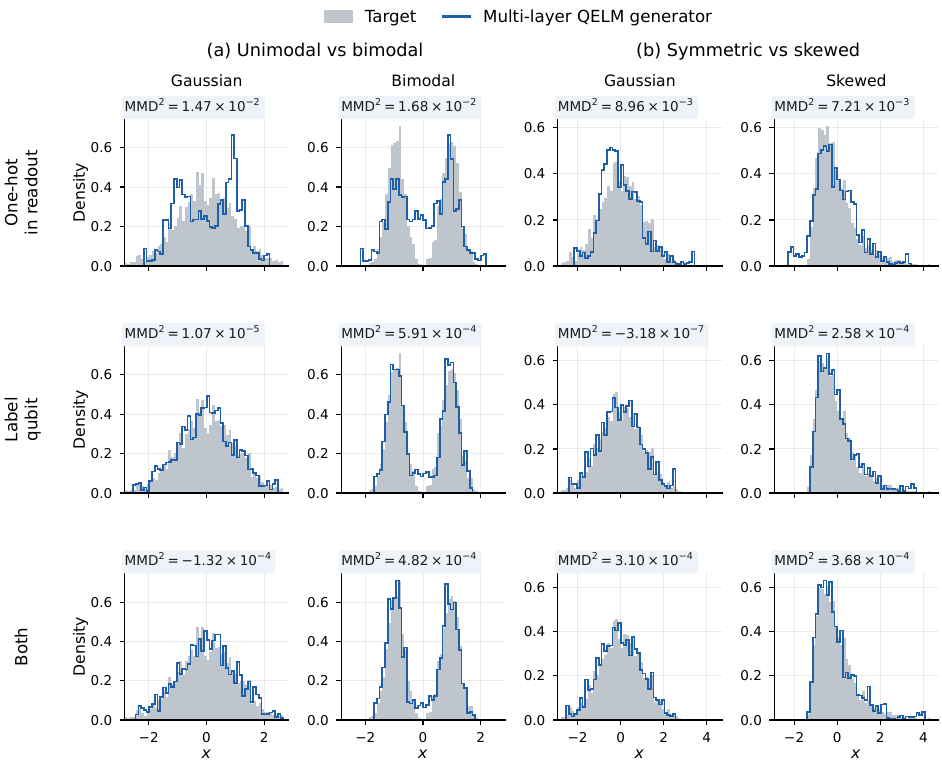}
  \caption{Two-class conditional generation in one dimension, for two target pairs whose classes have matched mean and variance and differ only in shape: (a)~Gaussian vs bimodal and (b)~Gaussian vs skewed (standardized log-normal). Rows show the three conditioning routes: the label as a one-hot vector in the readout only (all $N=6$ qubits carry the latent), the label on $N_c=1$ label qubit ($\omega_c\in\{0,\pi\}$, five latent qubits), and both. Each readout is trained for 6000 Adam steps with 256 samples per class per mini-batch and 2500 training latents per class, using ideal expectation values. Gray histograms show 3000 target samples and blue outlines 3000 generated samples per class, from the median of three training seeds. Each panel's MMD$^2$ is the evaluation energy distance for that class, averaged over the three seeds, the estimator is unbiased, so values near zero can be slightly negative.}

  \label{res:1d_cond_2}
\end{figure}

\begin{figure}[htbp]
  \centering
  \includegraphics[width=\columnwidth]{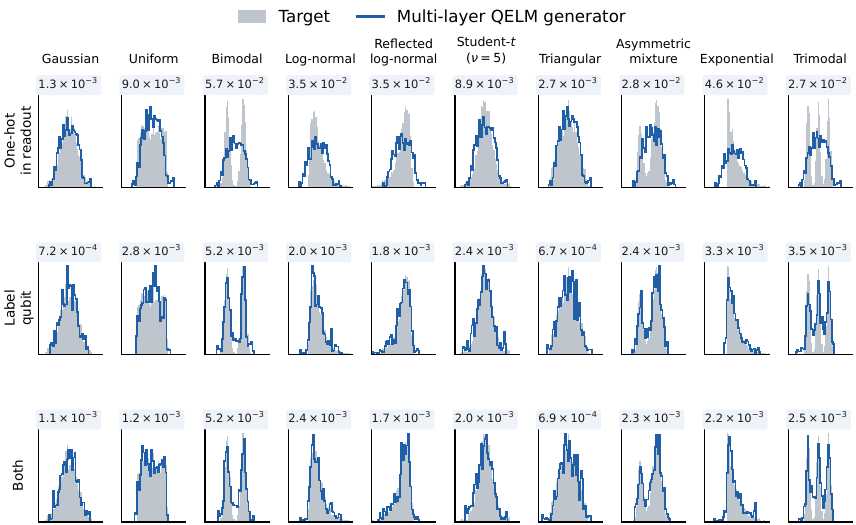}
  \caption{Ten-class conditional generation in one dimension. Each class is a different shape, standardized to zero mean and unit variance. Rows show the three conditioning routes as in Fig.~\ref{res:1d_cond_2}, with one label qubit the classes are encoded by the ten angles $\omega_c=\pi c/9$. Training as in Fig.~\ref{res:1d_cond_2}, with 1000 training latents per class. Gray histograms show the target and blue outlines the generated samples (3000 per class, median of three seeds). The number above each panel is that class's evaluation MMD$^2$, averaged over three seeds.}

  \label{res:1d_cond_10}
\end{figure}

Fig.~\ref{res:1d_cond_2} and Fig.~\ref{res:1d_cond_10} compare the three conditioning routes of Sec.~\ref{sec:cond_pred} on targets whose classes share the same mean and variance and differ only in shape, so that no per-class translation can tell them apart. With the label only in the readout, the generator produces essentially the same distribution for every class, a compromise between the class shapes: in Fig.~\ref{res:1d_cond_2}(a) the Gaussian and the bimodal class are generated as the same broad distribution, with humps where the bimodal modes lie and too much mass between them. Measured by the energy distance between the generated classes, divided by the same quantity for the target classes, this route separates the classes by $7.8\times10^{-4}$ (Gaussian vs bimodal) and $0.20$ (Gaussian vs skewed). This is the limitation derived in Sec.~\ref{sec:cond_pred}: a one-hot input can only translate a distribution, and these classes have the same mean.

Encoding the label on a single qubit removes this limitation. The generated classes are $0.93$ and $0.81$ times as far apart as the target classes, and the evaluation $\widehat{\mathrm{MMD}}^2$ falls by factors of $52$ and $63$, to $3.0\times10^{-4}$ and $1.3\times10^{-4}$. These values are comparable to those obtained between two independent sets of target samples ($4.2\times10^{-4}$ and $5.5\times10^{-4}$), so the remaining error is at the level of the sampling noise. Adding the one-hot vector on top of the label qubit gives no further improvement: the two routes agree within their seed-to-seed spread.

With ten classes on the same single label qubit (Fig.~\ref{res:1d_cond_10}), the label-qubit route is still ten times better than the one-hot route ($2.5\times10^{-3}$ vs $2.5\times10^{-2}$, averaged over classes), and it reproduces the bimodal, trimodal and asymmetric classes that the one-hot route blurs into a single unimodal shape. Its class separation, however, falls to $0.77$, as ten evenly spaced angles on one qubit must now carry all the class information through the same linear readout.

\subsection{Two-dimensional conditional target distributions}
\label{sec:2d_cond}

\begin{figure}[htbp]
  \centering
  \includegraphics[width=\columnwidth]{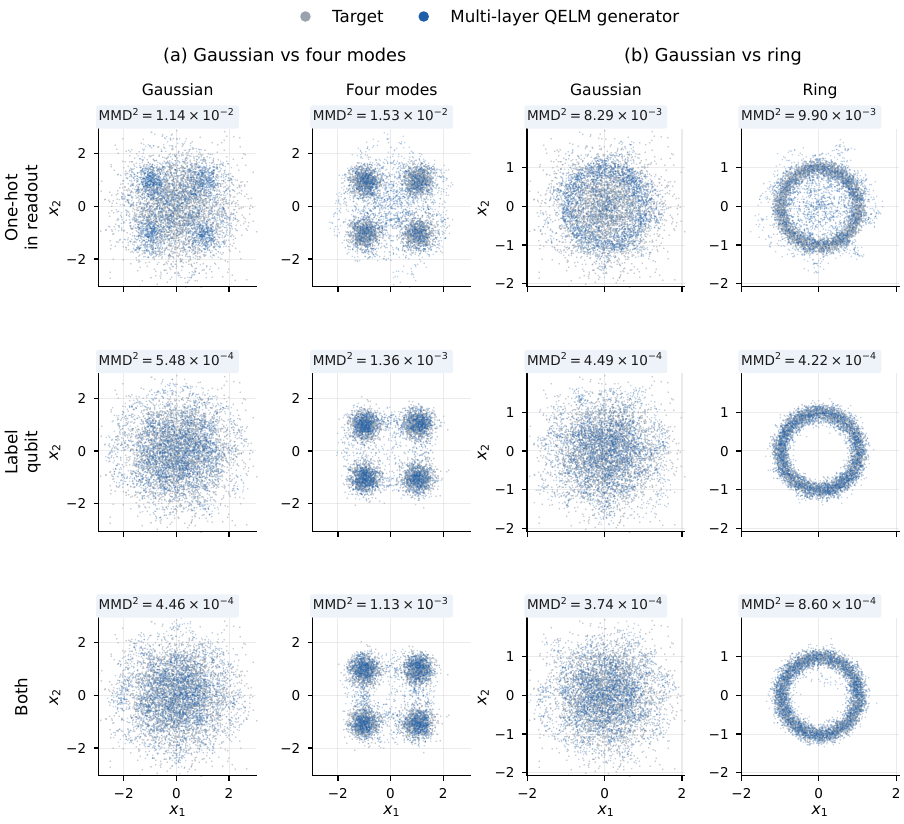}
  \caption{Two-class conditional generation in two dimensions, for class pairs with matched mean and covariance: (a)~Gaussian vs four modes and (b)~Gaussian vs ring. Rows show the three conditioning routes as in Fig.~\ref{res:1d_cond_2}. Every expectation value is estimated from 1000 measurement shots, with the shot noise drawn independently for every mini-batch and every generated sample (see Fig.~\ref{res:2d_support}), training is otherwise as in Fig.~\ref{res:1d_cond_2}. Each panel shows 3000 target samples (gray) and 3000 generated samples (blue) of one class, from the median of three seeds. Each panel's MMD$^2$ is that class's evaluation energy distance, averaged over three seeds.}

  \label{res:2d_cond_2}
\end{figure}

\begin{figure}[htbp]
  \centering
  \includegraphics[width=\columnwidth]{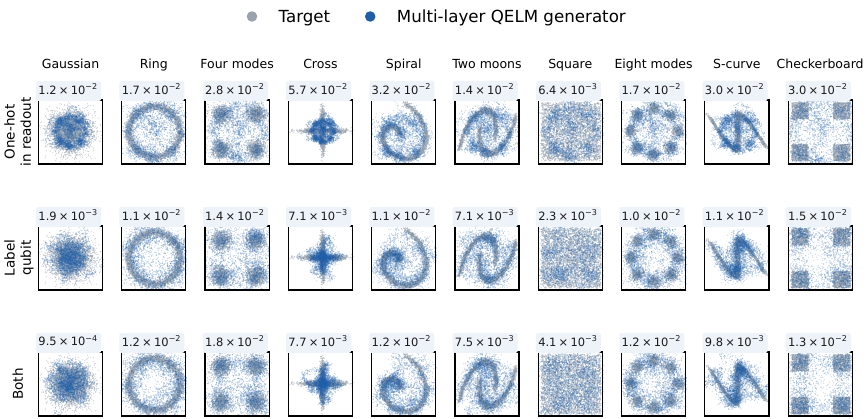}
  \caption{Ten-class conditional generation in two dimensions. Each class is a different shape, standardized to zero mean and identity covariance. Rows show the three conditioning routes as in Fig.~\ref{res:1d_cond_2}, with the ten classes encoded on one label qubit by $\omega_c=\pi c/9$. Every expectation value is estimated from 1000 measurement shots, and training uses 1000 latents per class. Each panel shows 3000 target samples (gray) and 3000 generated samples (blue) of one class, from the median of three seeds. The number above each panel is that class's evaluation MMD$^2$, averaged over three seeds.}

  \label{res:2d_cond_10}
\end{figure}

The same comparison in two dimensions is shown in Figs.~\ref{res:2d_cond_2} and~\ref{res:2d_cond_10}. Here every expectation value is estimated from 1000 measurement shots, which, as in Sec.~\ref{sec:results-2d}, spreads the generated samples off the one-dimensional curve and lets them fill the two-dimensional support. The pattern is the same as in one dimension. With the label only in the readout, the generator again produces one distribution for both classes, a superposition of the two target shapes, and the class separation is zero within the noise of the estimator for both target pairs. With one label qubit, the four separated modes and the ring are reproduced, with class separations of $0.85$ and $0.99$, and the evaluation MMD$^2$ is $14$ and $21$ times lower than with the one-hot route. Adding the one-hot vector again does not improve the evaluation MMD$^2$ beyond the seed-to-seed spread.

Ten two-dimensional classes (Fig.~\ref{res:2d_cond_10}) are the hardest case. The label-qubit route still outperforms the one-hot route, with an evaluation MMD$^2$ of $9.1\times10^{-3}$ against $2.4\times10^{-2}$ and a class separation of $0.46$ against $5.6\times10^{-3}$, and it captures the coarse structure of most classes, such as the ring, the four and eight modes, the cross and the checkerboard. Classes whose support is a thin curve, such as the spiral, the two moons and the S-curve, are only partly recovered: the generated samples follow the overall shape but spread across it. The current model reaches its limit when all ten classes are encoded on a single qubit and read out linearly.

\subsection{High-dimensional conditional target distributions}
\label{sec:multi-dim-res}

We next test the model on targets of much higher dimension: class-conditional generation of handwritten digits. We use two datasets. The first, Digits, contains $1797$ grayscale images of $8\times8$ pixels with 17 gray levels, so each class is a distribution on $\mathbb{R}^{64}$. The second, MNIST, contains $28\times28$ images, so each class is a distribution on $\mathbb{R}^{784}$. We use a random subset of 8000 images. In both datasets the ten classes are the digits $0$ to $9$, and pixel values are scaled to $[0,1]$. Each dataset is split into $70\%$ training and $30\%$ test images (1257/540 for Digits, 5600/2400 for MNIST).

\begin{figure}[htbp]
    \centering
    \includegraphics[width=\linewidth]{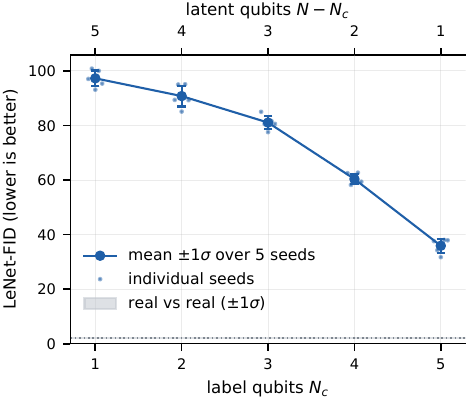}
    \caption{LeNet-FID of the MNIST generator as a function of the number of label qubits $N_c$, with the unbounded linear readout. The register has $N=6$ qubits, so the remaining $N-N_c$ qubits carry the latent $x$ (top axis); all label qubits receive the same angle $\omega_c=\pi c/9$. All runs use the Haar-random reservoir with $g=5$ and $L=7$ ($d=119$), the label-qubit route and the pixel-space plus LeNet-feature $\widehat{\mathrm{MMD}}^2$, trained for 3000 Adam steps. Large markers show the mean and standard deviation over 5 training seeds, and small dots the individual seeds. The gray band is the FID between two disjoint sets of real images of the same size ($\pm1\sigma$).}
    \label{res:nlabel}
\end{figure}

The reservoir is the same as in the previous sections ($N=6$, $g=5$, $L=7$, Haar-random $U$, $d=119$), and the label enters through label qubits (Sec.~\ref{sec:cond_pred}), now with $N_c=5$ (see Fig.~\ref{res:nlabel} for this choice): all five label qubits receive the same angle $\omega_c=\pi c/9$, and the remaining qubit carries the latent $x$. The readout $\mathbf{y}=W\mathbf{h}(x,c)+\mathbf{b}$ maps the $d=119$ features to all pixels at once, with $P(d+1)$ trained parameters for $P$ pixels: $7680$ for Digits and $94\,080$ for MNIST. For every class we draw 1000 latents and precompute their features, using 700 for training and 300 for evaluation.

In pixel space, the energy distance compares images through Euclidean distances between pixel vectors, which are a poor measure of whether two images show the same digit. We therefore add a perceptual term: the same $\widehat{\mathrm{MMD}}^2$ with the Riesz kernel (Eq.~\ref{eq:mmd_w_kernel}), evaluated on the 128-dimensional penultimate-layer features of a small convolutional classifier (LeNet) trained on MNIST and then frozen. Digits images are upsampled bilinearly to $28\times28$ so that the same MNIST-trained network can be applied to them. The per-class loss is
\begin{equation}
  \mathcal{L}_c = \widehat{\mathrm{MMD}}^2_{\mathrm{pix}} + \lambda\, s\, \widehat{\mathrm{MMD}}^2_{\mathrm{feat}},
\end{equation}
where $s$ is the ratio of the two terms, recomputed at every step and excluded from the gradient, so that $\lambda=0.5$ fixes the relative weight of the feature term regardless of the two terms' scales. The gradient flows through the frozen network to the generated pixels and from there to $(W,\mathbf{b})$. The feature term is essential. We measure this with the Fréchet distance between Gaussians fitted to the LeNet features of real and generated images~\cite{heusel2017gans}, which we call the LeNet-FID,
\begin{equation}
  \mathrm{FID} = \lVert \boldsymbol{\mu}_{\mathrm{r}} - \boldsymbol{\mu}_{\mathrm{g}} \rVert^2 + \mathrm{Tr}\big(\Sigma_{\mathrm{r}} + \Sigma_{\mathrm{g}} - 2\,(\Sigma_{\mathrm{r}}\Sigma_{\mathrm{g}})^{1/2}\big),
\end{equation}
where $(\boldsymbol{\mu}_{\mathrm{r}},\Sigma_{\mathrm{r}})$ and $(\boldsymbol{\mu}_{\mathrm{g}},\Sigma_{\mathrm{g}})$ are the mean and covariance of the 128-dimensional LeNet features of the real and generated images. The LeNet-FID is zero for identical Gaussians and grows as the two distributions move apart. Removing the feature term raises it from $5.6$ to $8.0$ on Digits and from $35.9$ to $92.8$ on MNIST. Each step uses 25 generated and 25 real images per class, and the readout is trained for 3000 Adam steps with learning rate $10^{-2}$.

We evaluate all generated images by the LeNet-FID. For every test image we generate one image conditioned on its class and compare the two sets. Pixels are clamped to $[0,1]$ before entering the network. As a reference, the FID between two disjoint sets of real images, each the size of the test set (540 images for Digits and 2400 for MNIST), is $2.69\pm0.74$ for Digits and $2.13\pm0.43$ for MNIST. These values are a finite-sample reference rather than a lower bound: FID is biased upward at finite sample size, so even two sets drawn from the same distribution give a nonzero value. All FIDs are averaged over 5 training seeds. Since the same LeNet features enter the training loss through $\widehat{\mathrm{MMD}}^2_{\mathrm{feat}}$, the LeNet-FID is not an independent measure of image quality; an evaluator not used during training, together with a measure of within-class diversity, would give a stricter assessment.

\begin{figure}[htbp]
  \centering
  \includegraphics[width=\columnwidth]{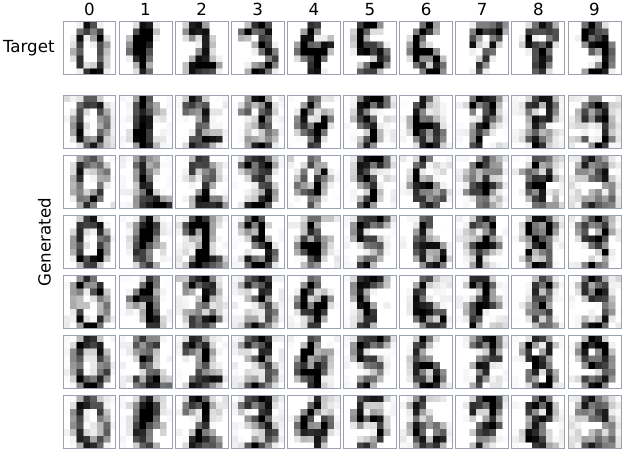}\\[2pt]
  {\small (a)~Digits $8\times8$}\\[6pt]
  \includegraphics[width=\columnwidth]{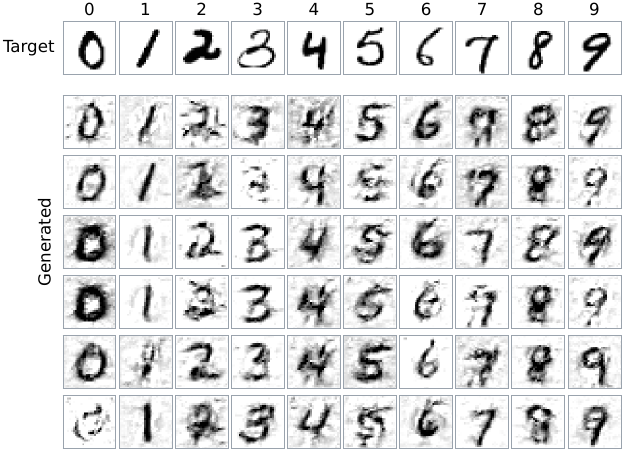}\\[2pt]
  {\small (b)~MNIST $28\times28$}
  \caption{Class-conditional image generation with the multi-layer QELM and an unbounded linear readout, for (a)~Digits and (b)~MNIST. The reservoir is Haar-random with $N=6$, $g=5$, $L=7$ ($d=119$), five label qubits carry $\omega_c=\pi c/9$ and one qubit carries the latent. The readout is trained for 3000 Adam steps on the pixel-space plus LeNet-feature $\widehat{\mathrm{MMD}}^2$. The top row shows one real test image per class, the six rows below show generated images for the first six test images of that class, with no selection. Pixel values are clipped to $[0,1]$ for display only. Images are from the training seed with the median FID, over 5 seeds the FID is $5.64\pm0.88$ (Digits) and $35.93\pm2.51$ (MNIST).}
  \label{res:images}
\end{figure}

\begin{figure}[htbp]
  \centering
  \includegraphics[width=\columnwidth]{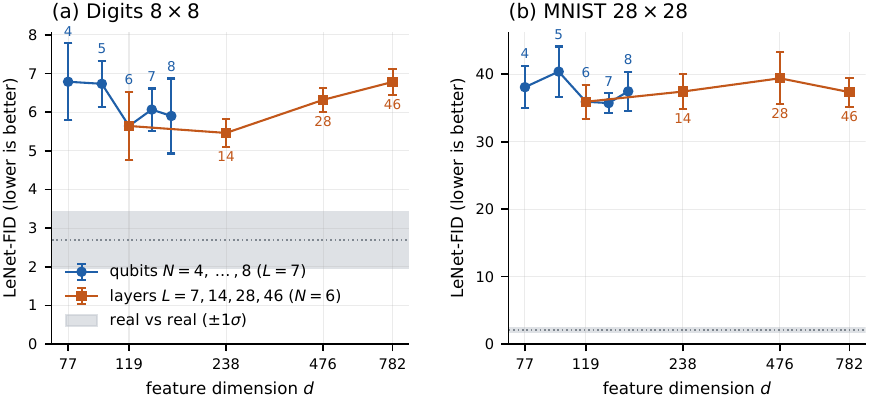}
  \caption{LeNet-FID of the image generator as a function of the feature dimension $d$, for (a)~Digits and (b)~MNIST, with an unbounded linear readout. Blue: $N=4,\dots,8$ qubits at $L=7$ layers, with $N_c=N-1$ label qubits and one latent qubit. Orange: $L=7,14,28,46$ layers at $N=6$. All runs use the Haar-random reservoir and the label-qubit route. Points show the mean and standard deviation over 5 training seeds. The gray band is the FID between two disjoint sets of real images of the same size ($\pm1\sigma$).}
  \label{res:scaling}
\end{figure}

Fig.~\ref{res:images} shows the generated images. On Digits (FID $5.64\pm0.88$) every generated image shows the digit it was conditioned on, with plausible variation in stroke thickness and slant. On MNIST (FID $35.93\pm2.51$) the classes are still recognizable, and a LeNet classifier assigns $98\%$ of the generated test images to the class they were conditioned on, but the strokes are blotchy and the background is gray instead of white. Fig.~\ref{res:scaling} shows that adding features does not remove this gap. Growing the reservoir from $N=4$ to $8$ qubits at $L=7$, or from $L=7$ to $46$ layers at $N=6$, changes $d$ tenfold, from $77$ to $782$, but the MNIST FID stays between $35.7$ and $40.4$ and the Digits FID between $5.5$ and $6.8$.

The choice of five label qubits matters. Fig.~\ref{res:nlabel} varies the number of label qubits $N_c$ at fixed $N=6$ on MNIST, with the remaining $N-N_c$ qubits carrying the latent. The FID falls monotonically from $97.3\pm2.9$ at $N_c=1$ to $35.9\pm2.5$ at $N_c=5$, and the fraction of generated test images that the LeNet classifier assigns to their intended class rises from $13\%$, close to chance, to $98\%$. With a single label qubit, the ten digit classes are therefore not separated, in contrast with the ten low-dimensional classes of Figs.~\ref{res:1d_cond_10} and~\ref{res:2d_cond_10}.

\begin{figure}[htbp]
  \centering
  \includegraphics[width=\columnwidth]{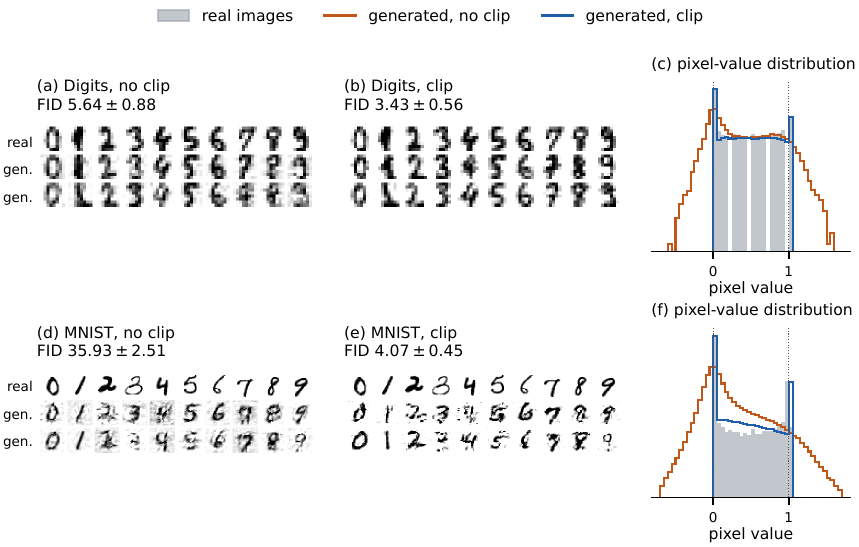}
  \caption{Effect of clipping the readout output, $\mathbf{y}=\mathrm{clip}(W\mathbf{h}+\mathbf{b},0,1)$, trained with a straight-through gradient. (a,~d)~Unbounded linear readout, (b,~e)~clipped readout, with everything else identical (Haar reservoir, $N=6$, $g=5$, $L=7$, $N_c=5$, pixel-space plus LeNet-feature $\widehat{\mathrm{MMD}}^2$, 3000 Adam steps). Each mosaic shows one real test image per class (top row) and generated images for the first two test images of that class, from the training seed with the median FID, pixel values are clipped to $[0,1]$ for display. The FID is the mean and standard deviation over 5 seeds, and pixels are clamped to $[0,1]$ before the FID evaluation in both cases. (c,~f)~Histogram of pixel values over all generated test images of the plotted seed, on a logarithmic density scale: real test images (gray), the unbounded readout output (orange) and the clipped output (blue). Dotted lines mark $0$ and $1$.}
  \label{res:clip}
\end{figure}

\textbf{Output clipping.} The readout output is unbounded, while pixel values lie in $[0,1]$. In the generated images of Fig.~\ref{res:images}, $29\%$ of Digits and $41\%$ of MNIST pixels fall outside $[0,1]$ (Fig.~\ref{res:clip}(c,~f)). The real images, by contrast, have a large exactly-white background: $49\%$ of Digits and $81\%$ of MNIST pixels are exactly zero. With a linear readout, the background value of every pixel is a linear combination of the same smooth features that draw the strokes, so the readout can keep it close to zero only on average.

We therefore clip the readout output, $\mathbf{y}=\mathrm{clip}(W\mathbf{h}(x,c)+\mathbf{b},\,0,\,1)$, element-wise in both training and generation. The clip is a fixed, parameter-free nonlinearity after the linear readout: the trained parameters are unchanged, the fixed reservoir still supplies all features, and the generator equals the linear readout wherever it lies in $[0,1]$. We train through the clip with a straight-through gradient~\cite{bengio2013estimating}, where the forward pass applies the clip and the backward pass treats it as the identity. The exact gradient of the clip is zero outside $[0,1]$, so pixels that start outside the range would never move back, and training with it fails (FID above $10^3$ on both datasets).

Fig.~\ref{res:clip} compares the two generators with everything else unchanged. Clipping lowers the FID from $5.64\pm0.88$ to $3.43\pm0.56$ on Digits and from $35.93\pm2.51$ to $4.07\pm0.45$ on MNIST, a ninefold reduction. Since the FID evaluation clamps pixels to $[0,1]$ for both generators, the gain comes from training through the clip, not from the range limit at evaluation. The pixel histograms show how. Trained through the clip, the readout drives background pixels far below zero, where the clip maps them to exactly zero: $78\%$ of generated MNIST pixels are exactly zero ($81\%$ for the real images), and $47\%$ on Digits ($49\%$). Without the clip, the same pixels are spread on both sides of zero. 

\begin{figure}[htbp]
  \centering
  \includegraphics[width=\columnwidth]{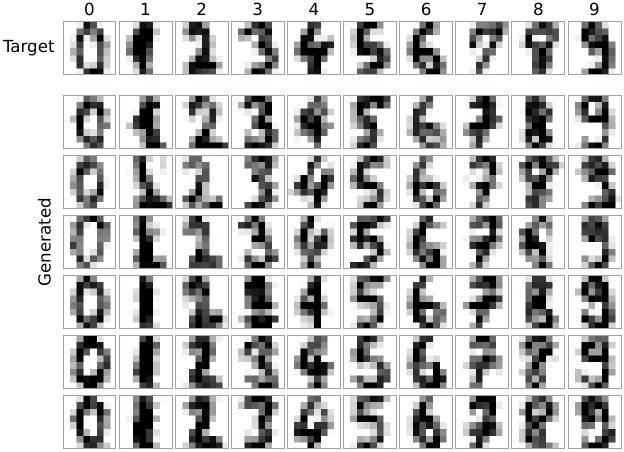}\\[2pt]
  {\small (a)~Digits $8\times8$}\\[6pt]
  \includegraphics[width=\columnwidth]{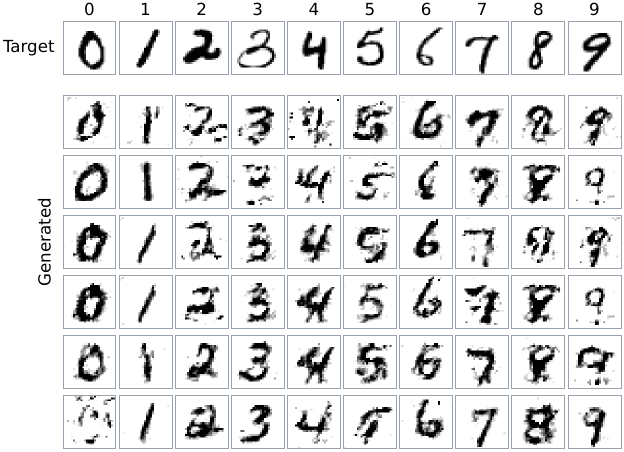}\\[2pt]
  {\small (b)~MNIST $28\times28$}
  \caption{Class-conditional image generation with the clipped readout, $\mathbf{y}=\mathrm{clip}(W\mathbf{h}(x,c)+\mathbf{b},0,1)$, otherwise as in Fig.~\ref{res:images}. Top row: one real test image per class, rows below: generated images for the first six test images of that class, with no selection, from the training seed with the median FID. Over 5 seeds the FID is $3.43\pm0.56$ (Digits) and $4.07\pm0.45$ (MNIST).}
  \label{res:images_clip}
\end{figure}

\begin{figure}[htbp]
  \centering
  \includegraphics[width=\columnwidth]{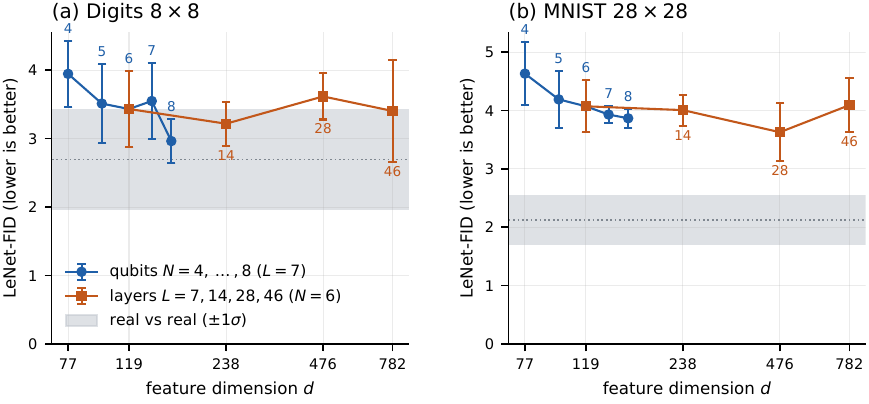}
  \caption{LeNet-FID as a function of the feature dimension $d$ with the clipped readout, for (a)~Digits and (b)~MNIST. The runs are those of Fig.~\ref{res:scaling} retrained with the output clip and a straight-through gradient, everything else is unchanged. Points show the mean and standard deviation over 5 training seeds. The gray band is the FID between two disjoint sets of real images of the same size ($\pm1\sigma$).}
  \label{res:scaling_clip}
\end{figure}

The clipped generator's samples are shown in Fig.~\ref{res:images_clip}: the MNIST background is now white and the strokes are sharper, although some samples still have broken or spurious strokes. Fig.~\ref{res:scaling_clip} repeats the size sweep of Fig.~\ref{res:scaling} with the clip. The FID is now $1.6$ to $2.0\times$ lower on Digits and $8$ to $11\times$ lower on MNIST in every cell, and trends that were previously hidden by the seed-to-seed spread become visible. Adding qubits lowers the FID on both datasets, from $4.63\pm0.54$ at $N=4$ to $3.87\pm0.16$ at $N=8$ on MNIST and from $3.94\pm0.48$ to $2.96\pm0.33$ on Digits. Adding layers has no consistent effect: from $L=7$ to $46$ the FID stays between $3.2$ and $3.6$ on Digits and between $3.6$ and $4.1$ on MNIST. For Digits, $N=8$ and $L=14$ lie within one standard deviation of the real-versus-real reference. MNIST remains $1.7$ to $2.2$ times above it.

\section{Choice of reservoir dynamics}
\label{sec:ham}

All results so far use a Haar-random unitary $U$. A Haar-random unitary on many qubits is not a natural operation for quantum hardware, so we now replace it by the time evolution $U=e^{-iHT}$ under a fixed Hamiltonian, as realized on an analog quantum simulator. We consider four families on $N$ qubits, all with random couplings drawn once and then held fixed: a mixed-field ring,
\begin{equation}\label{eq:mixed_field}
  \begin{aligned}
  H_{\mathrm{mix}} ={}& \sum_{i} \big( J^{z}_i\, Z_i Z_{i+1} + J^{x}_i\, X_i X_{i+1} \big)\\
    &+ \sum_{i} \big( h^z_i\, Z_i + h^x_i\, X_i \big),
  \end{aligned}
\end{equation}
a transverse-field Ising ring,
\begin{equation}\label{eq:trans_ising}
  H_{\mathrm{TFI}} = \sum_i J_i\, Z_iZ_{i+1} + \sum_i h_i\, X_i,
\end{equation}
a Heisenberg ring,
\begin{equation}\label{eq:heisenberg}
  \begin{aligned}
  H_{\mathrm{Heis}} ={}& \sum_i J_i \big( X_iX_{i+1} + Y_iY_{i+1} + Z_iZ_{i+1} \big)\\
    &+ \sum_i \big( h^z_i\, Z_i + h^x_i\, X_i \big),
  \end{aligned}
\end{equation}
and an all-to-all model with $ZZ$ and $XX$ couplings and local fields,
\begin{equation}\label{eq:all_to_all}
  \begin{aligned}
  H_{\mathrm{all}} ={}& \sum_{i<j} \big( J^z_{ij}\, Z_iZ_j + J^x_{ij}\, X_iX_j \big)\\
    &+ \sum_i \big( h^z_i\, Z_i + h^x_i\, X_i \big).
  \end{aligned}
\end{equation}
The rings have periodic boundary conditions. All couplings are drawn from $\mathcal{U}[0,1]$ and all fields from $\mathcal{U}[0.5,1]$, and the evolution time is $T=10$. For the mixed-field ring we also draw a second, independent set of couplings. Everything else is as in Sec.~\ref{sec:multi-dim-res}: the encoding gates and their offsets are identical across all reservoirs, so only $U$ changes.

To test whether the dynamics matter at all, we add two controls in which information spreads less or not at all: the identity, $U=I$, for which the qubits never interact and the latent and label qubits evolve independently, and the mixed-field ring evolved for shorter times, $T=0.25$, $1$ and $3$.

\begin{figure}[htbp]
  \centering
  \includegraphics[width=\columnwidth]{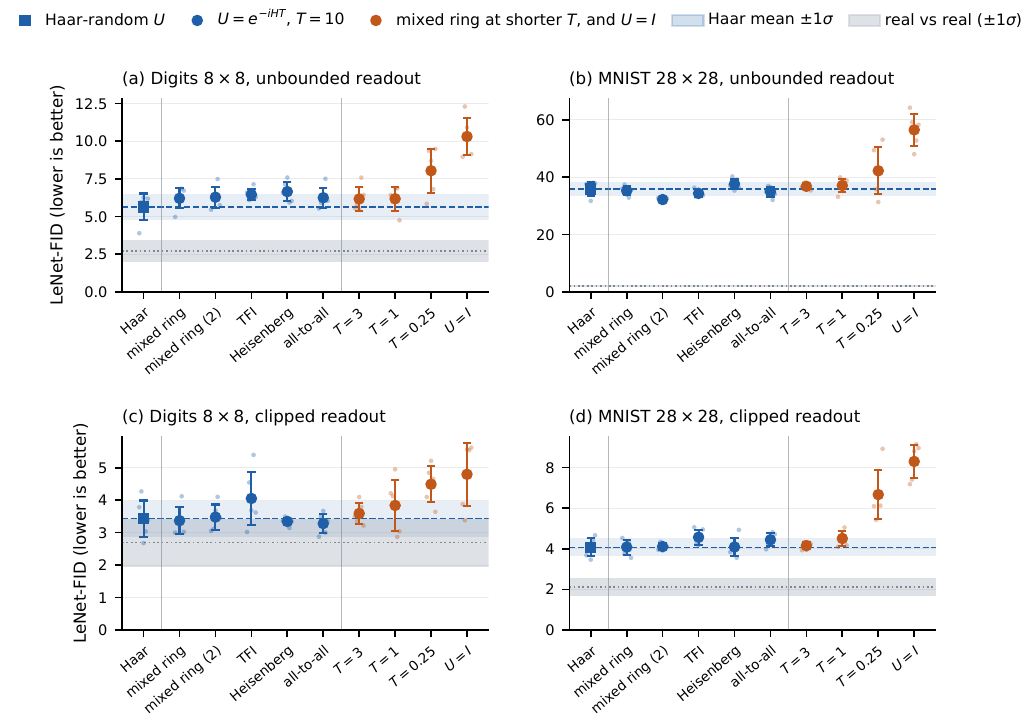}
  \caption{LeNet-FID of the image generator for different reservoir dynamics, for (a,~c)~Digits and (b,~d)~MNIST, with (a,~b)~the unbounded and (c,~d)~the clipped readout. The square is the Haar-random $U$ used in all other figures. Blue circles: $U=e^{-iHT}$ with $T=10$ for the mixed-field ring, Eq.~\eqref{eq:mixed_field} (two independent realizations of the couplings), the transverse-field Ising ring, Eq.~\eqref{eq:trans_ising}, the Heisenberg ring, Eq.~\eqref{eq:heisenberg}, and the all-to-all model, Eq.~\eqref{eq:all_to_all}. Orange: controls with less or no spreading of information, the mixed-field ring at $T=3$, $1$ and $0.25$, and $U=I$. All runs use $N=6$, $g=5$, $L=7$, $N_c=5$ label qubits and the pixel-space plus LeNet-feature $\widehat{\mathrm{MMD}}^2$; only $U$ changes. Large markers show the mean and standard deviation over 5 training seeds, and small dots the individual seeds. The dashed line and the blue band mark the Haar mean $\pm1\sigma$, and the gray band the FID between two disjoint sets of real images of the same size ($\pm1\sigma$).}
  \label{res:ham}
\end{figure}

Fig.~\ref{res:ham} shows the result. All five Hamiltonian reservoirs at $T=10$ perform like the Haar reservoir: over both datasets and both readouts their mean FID lies between $0.90$ and $1.18$ times the Haar value, and in 19 of the 20 comparisons the difference is within two standard errors. The one exception is the second realization of the mixed-field ring on MNIST with the unbounded readout, which is better than Haar ($32.2\pm0.6$ against $35.9\pm2.5$). Neither the form of the Hamiltonian, nor its connectivity (ring or all-to-all), nor the particular random couplings makes a consistent difference.

The controls show that, for this architecture, the dynamics nevertheless improve the generator substantially. Without any evolution, $U=I$, the register stays in a product state and the latent enters the features only through the single qubit that carries it. Latent and label still interact, since on the pair where they meet $\langle Z_iZ_{i+1}\rangle=\langle Z_i\rangle\langle Z_{i+1}\rangle$ is a product of a latent and a label function, but this gives only one such feature per layer. The FID then rises from $5.6$ to $10.3\pm1.2$ on Digits and from $35.9$ to $56.5\pm5.6$ on MNIST with the unbounded readout, and from $3.4$ to $4.8\pm1.0$ and from $4.1$ to $8.3\pm0.8$ with the clipped readout. Evolving the mixed-field ring for only $T=0.25$ gives intermediate values, while $T=1$ and $T=3$ are already within two standard errors of Haar in every case. The choice of Hamiltonian therefore does not matter. Within the dynamics we tested, what matters is that the evolution runs long enough to spread the latent and the label across the register before the observables are measured.

\section{Comparison with classical frameworks}
\label{sec:results-classical}

Since only a linear readout of fixed features is trained, the generator is a random-feature model, and the quantum reservoir enters only through the feature map $x\mapsto\mathbf{h}(x)$. The natural classical baseline therefore keeps everything else fixed (latents, readout, loss, optimizer and number of steps) and replaces $\mathbf{h}(x)$ by $M$ random Fourier features (RFF) of the same input,
\begin{equation}
  f_j(x) = \sqrt{2/M}\,\cos(w_j x + b_j), \qquad j=1,\dots,M,
\end{equation}
with frequencies $w_j\sim\mathcal{N}(0,\gamma^2)$ and phases $b_j\sim\mathcal{U}[0,2\pi]$ drawn once and then held fixed. The RFF map has two parameters, the number of features $M$ and the bandwidth $\gamma$. To compare feature maps of different size we use the effective rank $r_{\mathrm{eff}}=\exp\!\big(-\sum_i p_i\ln p_i\big)$, where $p_i=\lambda_i/\sum_k\lambda_k$ and $\lambda_i$ are the eigenvalues of the feature covariance over the latent samples.

\begin{figure*}[htbp]
  \centering
  \includegraphics[width=0.75\textwidth]{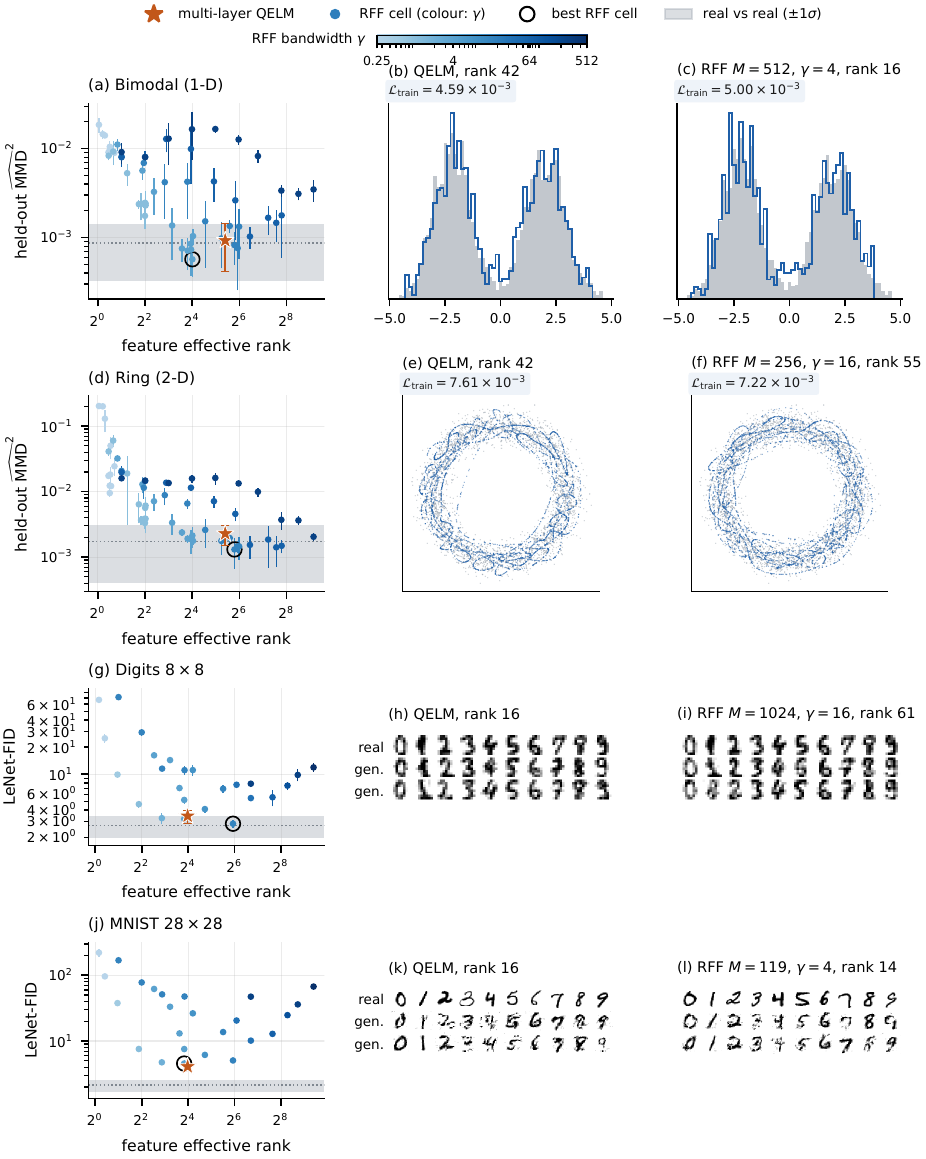}
  \caption{Multi-layer QELM (star) against random Fourier features (dots, colored by bandwidth $\gamma$) with the same linear readout, as a function of the feature effective rank, for (a)~the bimodal and (d)~the ring target and (g)~Digits and (j)~MNIST. Distribution rows: the protocol of Figs.~\ref{res:1d} and~\ref{res:2d}, with the held-out value of $\widehat{\mathrm{MMD}}^2$ (2000 against 2000 samples) on the vertical axis, $M\in\{2,\dots,1024\}$ and $\gamma\in\{0.25,\dots,256\}$, two RFF draws and two seeds per cell, three seeds for the QELM. Image rows: LeNet-FID with the clipped readout, five seeds, one RFF draw per cell. The circled dot is the best RFF cell. The gray band is the real-versus-real reference ($\pm1\sigma$). (b,~c,~e,~f,~h,~i,~k,~l)~Samples from the QELM and from the best RFF cell.}
  \label{res:rank}
\end{figure*}

Fig.~\ref{res:rank} sweeps $M$ from $2$ to $1024$ and $\gamma$ from $0.25$ to $256$ for two distribution targets and both image datasets. For every target the RFF error is U-shaped in $r_{\mathrm{eff}}$: with too few effective features the readout cannot represent the target, and with too many it overfits the finite set of training latents. The multi-layer QELM lies at the bottom of this curve on all four targets. Its error is within one combined standard deviation of the best RFF configuration: $(9.3\pm5.1)\times10^{-4}$ against $(5.7\pm2.1)\times10^{-4}$ on the bimodal target, $(2.3\pm0.8)\times10^{-3}$ against $(1.3\pm0.7)\times10^{-3}$ on the ring, an FID of $3.43\pm0.56$ against $2.82\pm0.30$ on Digits and $4.07\pm0.45$ against $4.54\pm0.60$ on MNIST. The QELM uncertainties are seed-to-seed spreads of the readout training for a single fixed reservoir realization; they do not include the variation across reservoir realizations, whereas the RFF spreads in Table~\ref{tab:matched} also include independent feature draws. On the two distribution targets both models are within the real-versus-real band. The best RFF configuration changes from target to target ($M$ between $119$ and $1024$, $\gamma$ between $4$ and $16$), whereas the QELM uses the same fixed reservoir throughout.

We then compare the two feature maps at equal size on all twelve distribution targets of Figs.~\ref{res:1d} and~\ref{res:2d}, with $M=d=119$ RFF features and $\gamma$ chosen per target by the training loss. We measure the accuracy by the held-out $\widehat{\mathrm{MMD}}^2$, between 2000 samples generated from latents not used in training and 2000 fresh target samples, and compare it with the same estimator evaluated between two independent sets of target samples. At this sample size both models reach the level of the sampling noise: the held-out $\widehat{\mathrm{MMD}}^2$ of the QELM lies within one standard deviation of the real-versus-real value on 11 of the 12 targets, and that of RFF on 9. RFF is lower on 6 of the targets, the median ratio of the two is $0.98$, and no difference exceeds two standard deviations in either direction (Table~\ref{tab:matched}). Varying the number of re-uploading gates from $g=1$ to $g=20$ does not change this: for every $g$ the held-out $\widehat{\mathrm{MMD}}^2$ stays at the real-versus-real level, and $g=1$ and $g=10$ differ by more than two standard deviations on only one target.

The same picture holds for conditional generation. With RFF features computed from the same inputs as the reservoir (the latent and the label angle) and $M=119$, the label route reaches the same evaluation MMD$^2$ as the QELM on all six conditional targets of Figs.~\ref{res:1d_cond_2}--\ref{res:2d_cond_10}, with no difference beyond two standard deviations (Table~\ref{tab:cond}, the table uses ideal features, so its values differ from the 1000-shot values of Sec.~\ref{sec:2d_cond}). The one-hot route fails in the same way for both feature maps, with class separations between $4.9 \times 10^{-4}$ and $0.22$. This confirms that the translation-only limit of Sec.~\ref{sec:cond_pred} is a property of the linear readout, not of the feature map.

Within the accuracy of these benchmarks, the multi-layer QELM therefore performs as well as a tuned classical random-feature model of the same size, but not better.

\begin{table}[htbp]
  \centering
  \caption{Held-out $\widehat{\mathrm{MMD}}^2$ ($\times10^{-3}$) of the multi-layer QELM and of $M=d=119$ random Fourier features on the twelve distribution targets of Figs.~\ref{res:1d} and~\ref{res:2d}, between 2000 generated and 2000 fresh target samples. The RFF bandwidth $\gamma$ is chosen per target by the training loss. Real vs real: the same estimator between two independent sets of target samples. Mean and standard deviation over three seeds (QELM) and two RFF draws times three seeds (RFF).}
  \label{tab:matched}
  \resizebox{\columnwidth}{!}{%
  \begin{tabular}{lccccc}
    \toprule
    Target & QELM & RFF & $\gamma$ & Real vs real & RFF/QELM \\
    \midrule
    Gaussian & $0.98\pm0.42$ & $0.71\pm0.27$ & $32$ & $1.27\pm1.07$ & $0.72$ \\
    log-normal & $0.67\pm0.23$ & $0.80\pm0.38$ & $64$ & $1.24\pm1.07$ & $1.19$ \\
    Student-$t$ & $1.36\pm0.93$ & $1.91\pm0.72$ & $64$ & $0.71\pm0.39$ & $1.40$ \\
    bimodal & $0.93\pm0.51$ & $1.53\pm0.47$ & $32$ & $0.87\pm0.55$ & $1.66$ \\
    skewed mixture & $1.41\pm1.10$ & $1.46\pm0.68$ & $32$ & $1.10\pm1.02$ & $1.03$ \\
    peak + exp.\ tail & $1.01\pm0.24$ & $1.16\pm0.79$ & $32$ & $0.92\pm0.56$ & $1.15$ \\
    corr.\ Gaussian & $2.02\pm0.44$ & $1.23\pm0.50$ & $128$ & $2.11\pm1.55$ & $0.61$ \\
    anti-corr.\ Gaussian & $1.35\pm0.12$ & $1.15\pm0.29$ & $128$ & $2.24\pm1.68$ & $0.85$ \\
    banana & $2.20\pm0.62$ & $1.76\pm0.74$ & $32$ & $1.75\pm0.98$ & $0.80$ \\
    ring & $2.28\pm0.79$ & $2.12\pm0.98$ & $32$ & $1.73\pm1.33$ & $0.93$ \\
    funnel & $2.07\pm0.82$ & $1.81\pm0.40$ & $256$ & $1.87\pm1.11$ & $0.87$ \\
    two moons & $1.54\pm0.43$ & $2.67\pm0.88$ & $16$ & $1.59\pm1.03$ & $1.73$ \\
    \bottomrule
  \end{tabular}}
\end{table}

\begin{table*}[htbp]
  \centering
  \caption{Conditional generation with the multi-layer QELM and with $M=119$ random Fourier features computed from the same inputs (latent and label angle), for the label route and the one-hot route. Evaluation MMD$^2$ ($\times10^{-3}$, averaged over classes) and class separation (energy distance between the generated classes divided by that between the target classes). In the one-hot route all classes are generated from the same latents, so the energy distance between generated classes is computed without the pairs that share a latent. Ideal features for all targets, mean over three seeds (two for ten classes).}
  \label{tab:cond}
  \small
  \begin{tabular}{lcccccccc}
    \toprule
    & \multicolumn{4}{c}{Label route} & \multicolumn{4}{c}{One-hot route} \\
    \cmidrule(lr){2-5}\cmidrule(lr){6-9}
    & \multicolumn{2}{c}{MMD$^2$} & \multicolumn{2}{c}{Separation} & \multicolumn{2}{c}{MMD$^2$} & \multicolumn{2}{c}{Separation} \\
    Targets & QELM & RFF & QELM & RFF & QELM & RFF & QELM & RFF \\
    \midrule
    1-D, Gaussian vs bimodal & $0.30$ & $0.44$ & $0.93$ & $0.92$ & $15.74$ & $15.30$ & $7.8\times10^{-4}$ & $4.9\times10^{-4}$ \\
    1-D, Gaussian vs skewed & $0.13$ & $0.03$ & $0.81$ & $0.99$ & $8.08$ & $7.81$ & $0.20$ & $0.22$ \\
    1-D, ten shapes & $2.60$ & $1.99$ & $0.77$ & $0.89$ & $24.78$ & $25.54$ & $0.07$ & $0.07$ \\
    2-D, Gaussian vs four modes & $1.42$ & $1.02$ & $0.97$ & $0.96$ & $14.26$ & $12.54$ & $3.6\times10^{-3}$ & $1.5\times10^{-3}$ \\
    2-D, Gaussian vs ring & $0.75$ & $0.51$ & $1.10$ & $1.06$ & $8.70$ & $7.72$ & $4.2\times10^{-3}$ & $3.0\times10^{-3}$ \\
    2-D, ten shapes & $6.94$ & $7.20$ & $0.55$ & $0.65$ & $25.68$ & $22.72$ & $5.4\times10^{-3}$ & $4.9\times10^{-3}$ \\
    \bottomrule
  \end{tabular}
\end{table*}

\section{Discussion and conclusions}
\label{sec:discussion}

We have shown that a fixed multi-layer QELM, of which only a single linear readout is trained, can serve as a generative model. Driven by a single scalar latent, it reproduces unimodal, multimodal, skewed and heavy-tailed one-dimensional distributions. Similarly, in two dimensions it is able to generate a variety of distributions, however, its samples lie on a curve, which finite-shot measurement noise thickens to fill the target's support. Encoding the class label on dedicated qubits enables conditional generation of classes that differ only in shape, which a one-hot label appended to the readout cannot achieve. 
On handwritten digits the generator reaches a LeNet-FID of $3.4$ on Digits and $4.1$ on MNIST once its output is clipped to the pixel range, compared with $2.7$ and $2.1$ between two sets of real images.

The comparisons of Secs.~\ref{sec:ham} and~\ref{sec:results-classical} clarify what the quantum reservoir contributes. The particular dynamics do not matter: four Hamiltonian families with different interactions and connectivity, and two random realizations of one of them, perform like a Haar-random unitary, provided the evolution runs long enough to spread the latent and the label across the register. The reservoir thus acts as a generic scrambling feature map. A classical random-feature map of the same size, with its bandwidth tuned, performs equally well on all targets we tested, and neither is better within the seed-to-seed spread. Since the generator is linear in its features, it is a random-feature model, and at the sizes studied here the quantum feature map offers no measurable advantage over a classical one.

While we were writing this work, P{\l}odzie{\'n}~\cite{plodzien2026generative} proposed a closely related generative QELM, with a fixed reservoir and a trained affine readout. These results agree with ours on two points: the particular reservoir unitary does not systematically improve the generator, and the quantum reservoir gives no advantage over its classical control. They differ on the role of the dynamics, especially for our conditional predictions: in their model, where every qubit carries a coordinate of the latent, a reservoir with no gates is competitive with entangling circuits, while in ours, where the latent and the label enter on separate qubits, removing the evolution degrades the generated images.

Whether the model improves with size remains an open question, and our results point towards unfavorable scaling. Growing the reservoir from $N=4$ to $8$ qubits, or from $L=7$ to $46$ layers, increases the number of features tenfold, from $d=77$ to $782$ (Figs.~\ref{res:scaling} and~\ref{res:scaling_clip}). With the unbounded readout this leaves the FID unchanged on both datasets. With the clipped readout, where the seed-to-seed spread is small enough to resolve trends, adding layers still gives no improvement, while adding qubits lowers the FID by $16$ to $25\%$ between $N=4$ and $8$. Part of this gain may come from the conditioning rather than from the feature space, since each added qubit is also an added label qubit ($N_c=N-1$), and at fixed $N$ the FID depends strongly on $N_c$ (Fig.~\ref{res:nlabel}).

There are two reasons to expect that this modest gain will not continue. First, more features are not always better. For the random Fourier features of Fig.~\ref{res:rank} the error is U-shaped in the effective rank: beyond an optimal value, additional features let the readout overfit the finite set of training latents, and the test error rises. The flat dependence on depth is consistent with the reservoir already being close to this optimum, and a larger reservoir, with features of higher effective rank, can be expected to move past it. Second, larger reservoirs suffer from exponential concentration. A scrambling unitary drives the reduced state of every qubit towards the maximally mixed state, so the local expectation values we measure vary less and less with the input, with typical fluctuations of order $2^{-N/2}$~\cite{xiong2023fundamental}. With exact expectation values this only rescales the features, which the linear readout can compensate. On hardware, however, each expectation value must be estimated from measurement shots, and resolving fluctuations of order $2^{-N/2}$ requires a number of shots that grows as $2^N$. Together, these effects suggest that simply enlarging the reservoir will not improve the generator, and that better performance will have to come from how the features are used rather than from their number.

Several limitations follow from the construction itself. A deterministic map of a scalar latent can only produce one-dimensional sets of samples, so any target of higher intrinsic dimension needs an additional source of randomness, such as the shot noise used here. The readout's output is a linear combination of bounded expectation values, which limits the range and tails it can reach, and on images it needed a fixed nonlinearity, the output clip, to produce an exactly white background. Finally, all results use ideal or simulated finite-shot expectation values. Running the model on quantum hardware, where shot noise is unavoidable and the dynamics are an analog Hamiltonian evolution, is the natural next step.

\bibliographystyle{apsrev4-2}
\bibliography{biblio}
\end{document}